\documentclass[11pt]{article}
\usepackage{moriond,epsfig}

\def\Journal#1#2#3#4{{#1} {\bf #2}, #3 (#4)}

\def\NIMA{{\em Nucl. Instrum. Methods} A}

\def\PLB{{\em Phys. Lett.}  B}
\def\PRL{\em Phys. Rev. Lett.}
\def\PRD{{\em Phys. Rev.} D}

\def\be{\begin{equation}}
\def\ee{\end{equation}}
\def\bea{\begin{eqnarray}}
\def\eea{\end{eqnarray}}

\begin{document}
\vspace*{4cm}
\title{Charmed baryon spectroscopy with Belle}

\author{ Tadeusz Lesiak \footnote{partially supported by the KBN grant No. 2P03B 01324} }

\address{Institute of Nuclear Physics PAN,
Radzikowskiego 152, 31-142 Krak\'{o}w, Poland}

\maketitle\abstracts{
Recent studies of charmed baryon spectroscopy, performed by the Belle
collaboration, are briefly described.
We report the observation of an isotriplet of baryons $\Sigma_c(2800)$ 
and evidence for two new baryons $\Xi_{cx}(2980)$ and  $\Xi_{cx}(3077)$.
Finally we present a precise
determination of the masses of $\Xi_c$ and $\Xi_c(2645)$. 
}

\section{Introduction}

In the last three years the Belle collaboration
has provided evidence
for several new hadrons. 
Among them are the states 
 X(3872), Y(3940) and Z(3931), 
discussed at this conference in a separate talk.~\cite{CHISTOV}
This paper is devoted to the recent studies by the Belle collaboration
concerning charmed baryons. In the first two chapters the observations of an isotriplet
of baryons $\Sigma_c(2800)$ and of two new states $\Xi_{cx}(2980)$ and
$\Xi_{cx}(3077)$ are briefly presented.
Chapters~\ref{XIC} and~\ref{CH_XC2645} are devoted to the precise mass determination
of the $\Xi_c$ and $\Xi_c(2645)$, respectively.
The Belle detector at the  KEKB asymmetric 
$e^+e^- $ collider~\cite{KEKB}
is a general purpose
spectrometer, described in detail in Ref.~\cite{BELLE}

\section{Observation of isotriplet of baryons  $\Sigma_c(2800)$}
\label{SIG2800}

\begin{figure}
\setlength{\unitlength}{1mm}
\begin{center}
\begin{picture}(230,60)(-12,0)
\put(-1,20){\rotatebox{90}{\tiny\bf events/(10~{\rm MeV}/$c^2$)}}
\put(43,1){{\large $M(\Lambda_c^+\pi)-M(\Lambda_c^+)$ [GeV/$c^2$]}}
\put(25,45){\large $\Lambda_c^+\pi^-$}
\put(63,45){\large $\Lambda_c^+\pi^0$}
\put(103,45){\large $\Lambda_c^+\pi^+$}
\centering\epsfig{figure=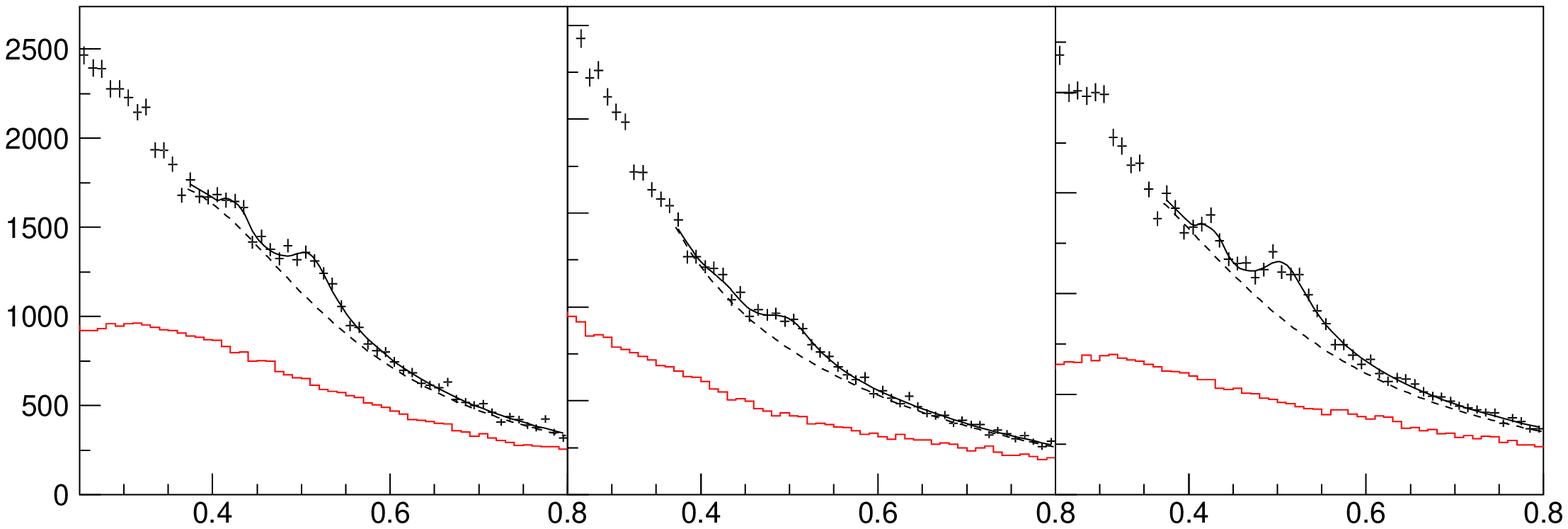,height=6cm}
\end{picture}
\end{center}
\caption{$M(\Lambda_c^+\pi)-M(\Lambda_c)$ distributions of the selected
$\Lambda_c^+\pi^-$ (left), $\Lambda_c^+\pi^0$ (middle), and $\Lambda_c^+\pi^+$ (right)
 combinations. Data from the $\Lambda_c^+$ signal window (points with error bars) and 
normalized sidebands (histograms) are shown, together with the fits (solid curves) and
 their combinatorial background components (dashed).}
\label{MIZUK}
\end{figure}

The Belle collaboration, using the data sample of 281 fb$^{-1}$, has provided 
the first evidence~\cite{MIZUK}
 for an isotriplet of excited charmed baryons $\Sigma_c(2800)$
decaying into the $\Lambda_c^+\pi^-$, $\Lambda_c^+\pi^0$ and $\Lambda_c^+\pi^+$ final 
states~\footnote{Charge-conjugate modes are included everywhere, unless otherwise stated}.
As shown in Fig.~\ref{MIZUK}, clear enhancements around 0.51 GeV/c$^2$ are seen in the
distributions of the mass difference 
$\Delta M (\Lambda_c^+\pi)= M(\Lambda_c^+\pi) - M(\Lambda_c^+)$
for the $\Lambda_c^+\pi^-$, $\Lambda_c^+\pi^0$, and $\Lambda_c^+\pi^+$ combinations.
The mass differences  $\Delta M$ together with the widths
of the states $\Sigma_c(2800)$ are collected in table~\ref{TABLE_MIZUK}.
These states are tentatively identified as the members of the predicted
$\Sigma_{c2}$, $J^P=3/2^-$ isospin triplet~\cite{MIZUK_THEOR}.
 The enhancement near $\Delta M=0.43$ GeV/c$^2$
(cf Fig.~\ref{MIZUK}),
in the spectra corresponding to 
$\Lambda_c^+\pi^-$ and 
$\Lambda_c^+\pi^+$ combinations, is attributed to feed-down from the decay 
$\Lambda_c(2880)^+\to \Lambda_c^+\pi^+\pi^-$, as verified by reconstructing
$\Lambda_c(2880)$ in the data.

\begin{table}[t]
\caption{Parameters of the baryons $\Sigma_c(2800)^0$, $\Sigma_c(2800)^+$ and $\Sigma_c(2800)^{++}$}
\vspace{0.4cm}
\begin{center}
\begin{tabular}{|l|c|c|c|c|}
\hline
State & $\Delta M$ [{\rm MeV}/c$^2$] & Width [{\rm MeV}] & Yield/$10^3$ & Significance ($\sigma$)   \\ 
\hline
$\Sigma_c(2800)^0$    &  $515.4^{+3.2+2.1}_{-3.1-6.0} $ & $61^{+18+22}_{-13-13}$ 
 & $2.24^{+0.79+1.03}_{-0.55-0.50}$ & ~8.6 \\
$\Sigma_c(2800)^+$    &  $~505.4^{+5.8+12.4}_{-4.6-2.0}$ & $62^{+37+52}_{-23-38}$ 
 & $1.54^{+1.05+1.40}_{-0.57-0.88}$ & ~6.2 \\
$\Sigma_c(2800)^{++}$ &  $514.5^{+3.4+2.8}_{-3.1-4.9} $ & $75^{+18+22}_{-13-11}$ 
 & $2.81^{+0.82+0.71}_{-0.60-0.49}$ & 10.0 \\
\hline
\end{tabular}
\end{center}
\label{TABLE_MIZUK}
\end{table}

\begin{figure}[th]
\setlength{\unitlength}{1mm}
\begin{picture}(190,50)
\put(13,44){{\large\bf (a) }}
\put(147,44){{\large\bf (b) }}
\begin{minipage}[b]{.5\linewidth}
\centering\epsfig{figure=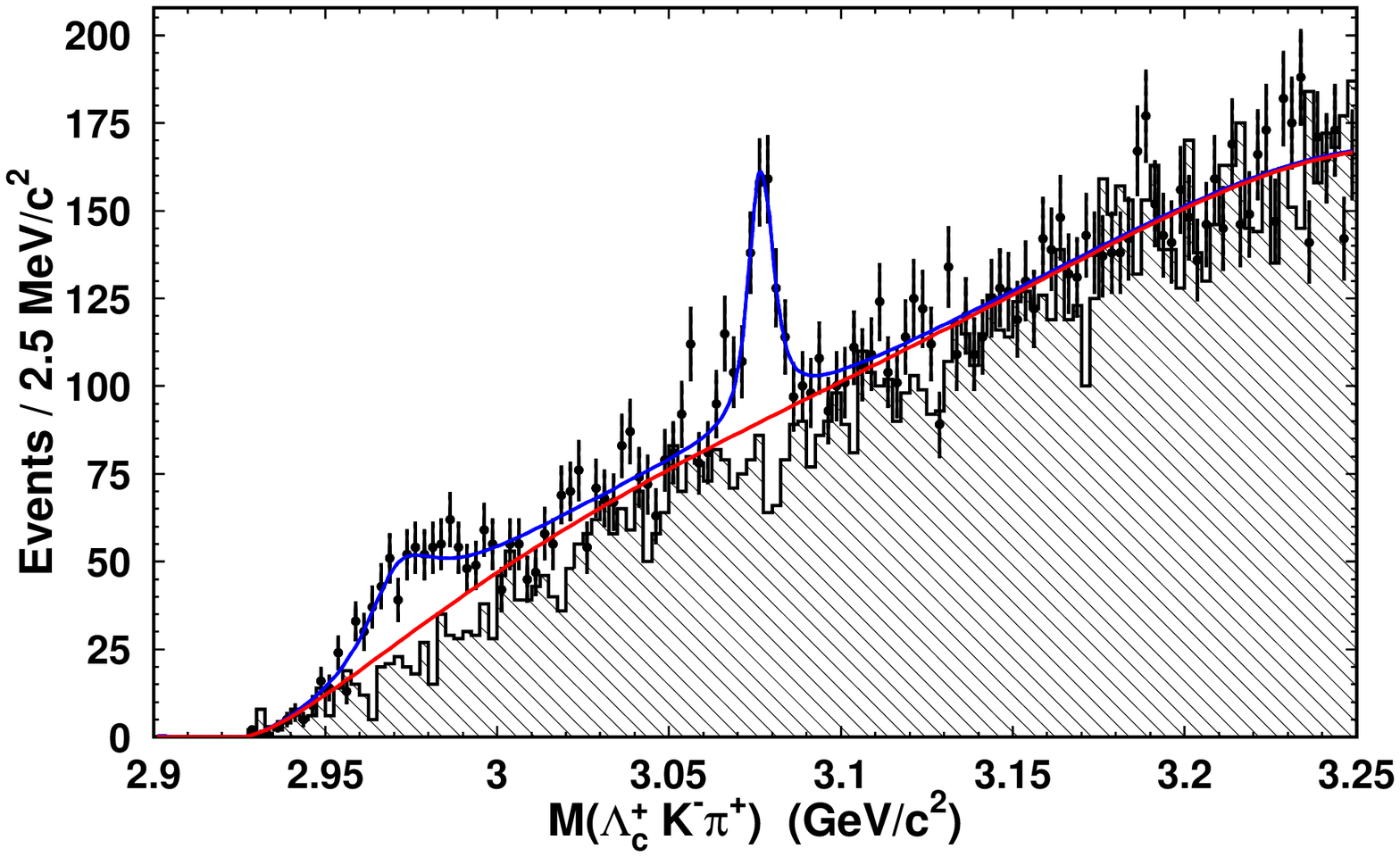,width=\linewidth,height=5.5cm}
\end{minipage}\hfill
\begin{minipage}[b]{.5\linewidth}
\centering\epsfig{figure=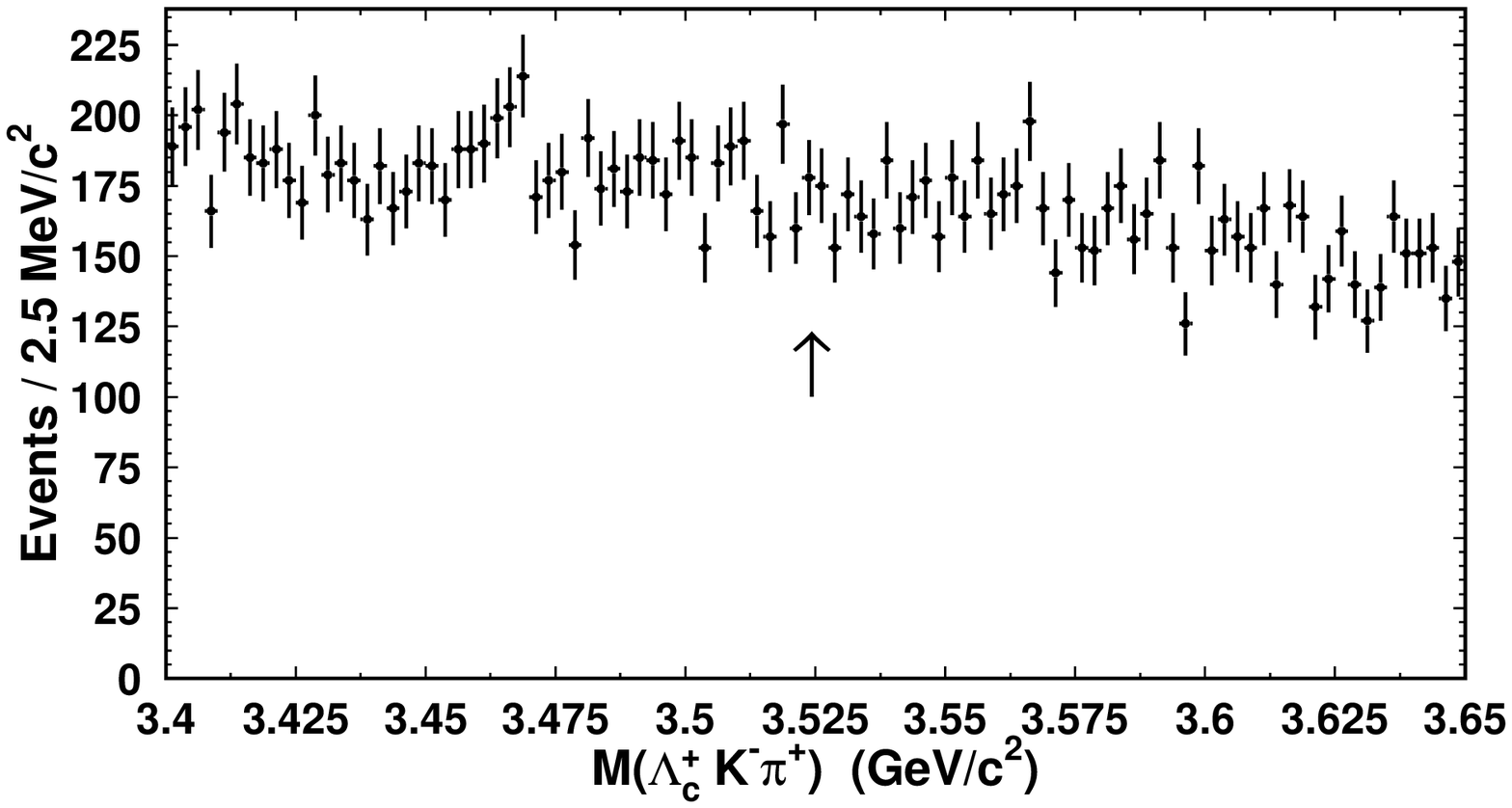,width=\linewidth,height=5.5cm}
\end{minipage}
\end{picture}
\caption{ {\bf (a)}: $M(\Lambda_c^+ K^-\pi^+)$ distribution (points with 
error bars) together with the fit (solid curve). The dashed region
represents the background component corresponding to the wrong-sign
combinations $\Lambda_c^+K^+\pi^-$. The signal of the $\Xi_{cx}(2980)^+$
is parametrized by a Breit-Wigner function while the one corresponding to the
$\Xi_{cx}(3077)^+$ is described by the convolution of a Breit-Wigner function
with the Gaussian detector resolution function.
{\bf (b)}: The distribution of $\Lambda_c^+ K^-\pi^+$ invariant mass
in the region around the value of 3.52 GeV (marked with an arrow)
corresponding to the SELEX evidence of the double-charmed baryon.
\label{CHISTOV}}
\end{figure}

\section{Observation of new states  $\Xi_{cx}(2980)$ and $\Xi_{cx}(3077)$}
\label{XICX}

\begin{table}
\caption{Parameters of the two new charm-strange baryons $\Xi_{cx}(2980)^+$ and
$\Xi_{cx}(3077)^+$}
\vspace{0.4cm}
\begin{center}
\begin{tabular}{|l|c|c|c|c|}
\hline
State & Mass [{\rm MeV/c}$^2$] & Width [{\rm MeV}] & Yield (events) & Significance [$\sigma$] \\ 
\hline
$\Xi_{cx}(2980)^+$  &  $2978.5\pm 2.1\pm 2.0$ & $43.5\pm 7.5\pm 7.0$ & $403.5\pm 50.7$ & 6.3 \\
$\Xi_{cx}(3077)^+$  &  $3076.7\pm 0.9\pm 0.5$ & $~6.2\pm 1.2\pm 0.8$ & $326.0\pm 39.6$ & 9.7 \\
\hline
\end{tabular}
\end{center}
\label{TABLE_CHISTOV}
\end{table}

At this Conference, the Belle collaboration using the data sample of 461.5 fb$^{-1}$, 
reported the first observation of two baryons, denoted as 
$\Xi_{cx}(2980)^+$ and
$\Xi_{cx}(3077)^+$ and 
decaying into $\Lambda_c^+ K^-\pi^+$ (Fig.~\ref{CHISTOV}(a)).
Possible reflections from decays of excited $\Lambda_c$ states
have been carefully examined using simulated and real data. Their contributions
are found to be negligible.
 Assuming that these
states carry charm and strangeness, the above observation 
would comprise the first example of a baryonic decay in which the initial $c$ and $s$ 
quarks are carried away by two different final state particles.
The  preliminary parameters of the states 
$\Xi_{cx}(2980)^+$ and
$\Xi_{cx}(3077)^+$ are collected in table~\ref{TABLE_CHISTOV}.
Most naturally, these two states would be interpreted as excited charm-strange
baryons $\Xi_c$. Before reaching the conclusion on the nature of the states
 $\Xi_{cx}(2980)^+$ and
$\Xi_{cx}(3077)^+$, further studies of their  properties are foreseen.

In the $\Lambda_c^+ K^-\pi^+$ final state, the SELEX collaboration~\cite{SELEX} reported
the observation of a double charmed baryon at 3520 MeV/c$^2$. The study by Belle 
shows no evidence for this state (Fig.~\ref{CHISTOV}(b), the quantitative determination of the 
production upper-limit is ongoing).

\section{Mass determination of the $\Xi_c$}
\label{XIC}

The Belle collaboration precisely determined the masses of the  hyperons
$\Xi_c^+$ and $\Xi_c^0$. This measurement~\cite{XIC} is based on the data sample of 140 fb$^{-1}$.
Around 5000 $\Xi_c^+$'s are reconstructed in three final states:
$\Xi^-\pi^+\pi^+$, $\Lambda K^-\pi^+\pi^+$ and  $p K^0_s K^0_s$ as well as
 8700  $\Xi_c^0$'s in the following  four final states:
$\Xi^-\pi^+$, $\Lambda K^-\pi^+$,  $\Lambda K^0_s$ and $p K^- K^- \pi^+$.

The average masses of the~$\Xi_c^0$ and~$\Xi_c^+$ have been determined to be 
\begin{equation} 
m_{\Xi_c^+}  = (2468.1\pm 0.4\,{\rm (stat. \oplus syst.)}^{+0.2}_{-1.4})\,{\rm MeV}/c^2 
\end{equation} 
\begin{equation} 
m_{\Xi_c^0}  = (2471.0\pm 0.3\,{\rm (stat. \oplus syst.)}^{+0.2}_{-1.4})\,{\rm MeV}/c^2, 
\end{equation} 
where the first error is the combined statistical and systematic uncertainty,
and the second is the uncertainty due to possible biasses in the mass scale.
We therefore find the $\Xi_c^0 - \Xi_c^+$ mass splitting to be
\begin{equation} 
m_{\Xi_c^0} - m_{\Xi_c^+} = (2.9\pm 0.5)\,{\rm MeV}/c^2. 
\end{equation} 
In addition, for the above mentioned decays, the branching ratios with respect
to the decays 
$\Xi_c^+\to \Xi^-\pi^+\pi^+$ and $\Xi_c^0\to \Xi^-\pi^+$ 
are determined~\cite{XIC}.

\section{Mass determination of the $\Xi_c(2645)$}
\label{CH_XC2645}

\begin{figure}[hbt]
\setlength{\unitlength}{1mm}
\begin{picture}(190,50)
\put(-3,24){\rotatebox{90}{\tiny\bf events/(5~{\rm MeV}/$c^2$)}}
\put(62,-2){{\large $M(\Xi_c\pi)$ [GeV/$c^2$]}}
\put(17,40){{\large $\Xi_c^+\pi^-$ }}
\put(17,35){{\large $\Xi_c^+\to \Xi^-\pi^+\pi^+$}}
\put(72,40){{\large $\Xi_c^0\pi^+$ }}
\put(72,35){{\large $\Xi_c^0\to \Xi^-\pi^+$}}
\put(125,40){{\large $\Xi_c^0\pi^+$ }}
\put(125,35){{\large $\Xi_c^0\to \Lambda^0 K^-\pi^+$}}
\begin{minipage}[b]{.33\linewidth}
\centering\epsfig{figure=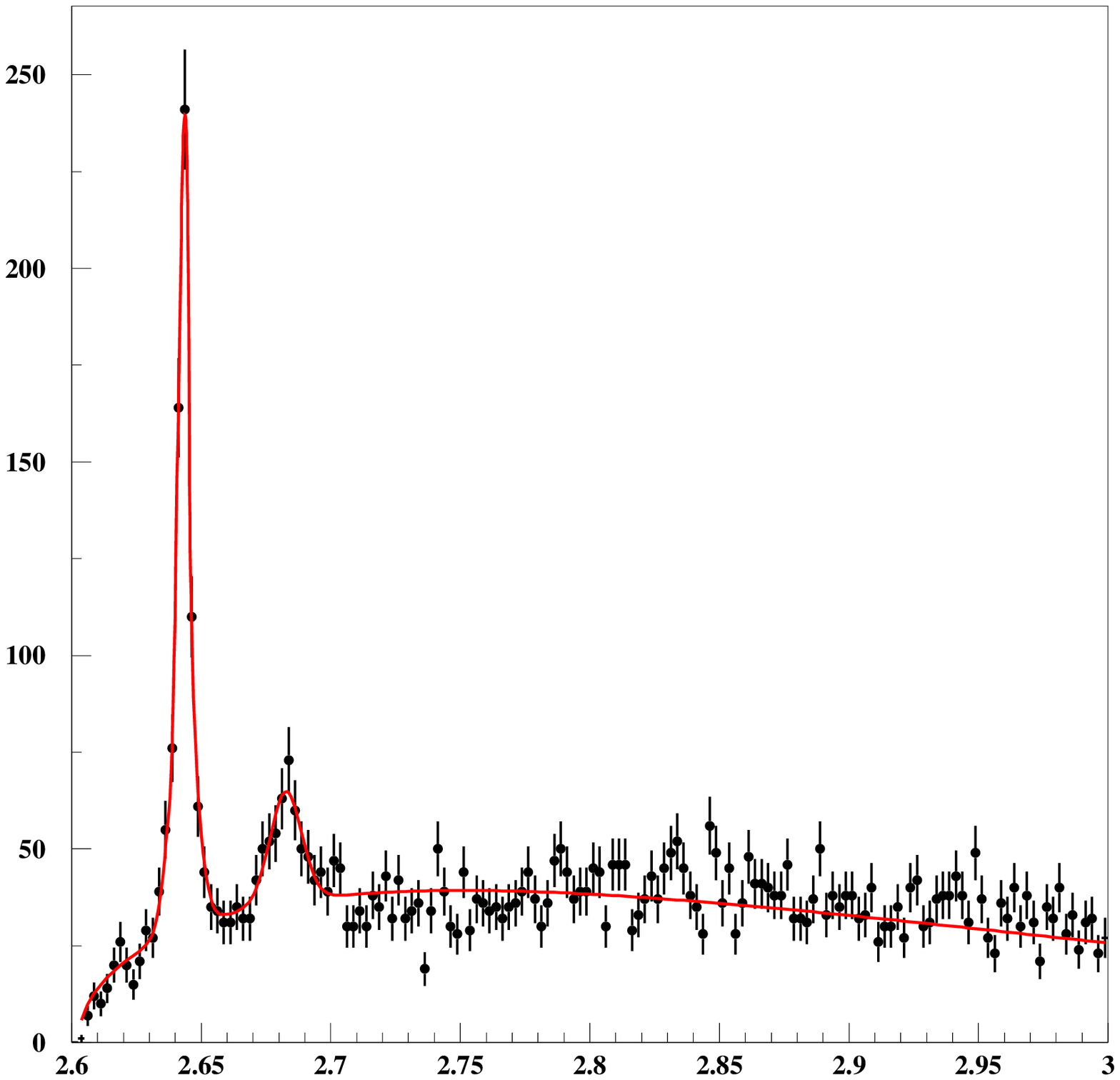,width=\linewidth}
\end{minipage}\hfill
\begin{minipage}[b]{.33\linewidth}
\centering\epsfig{figure=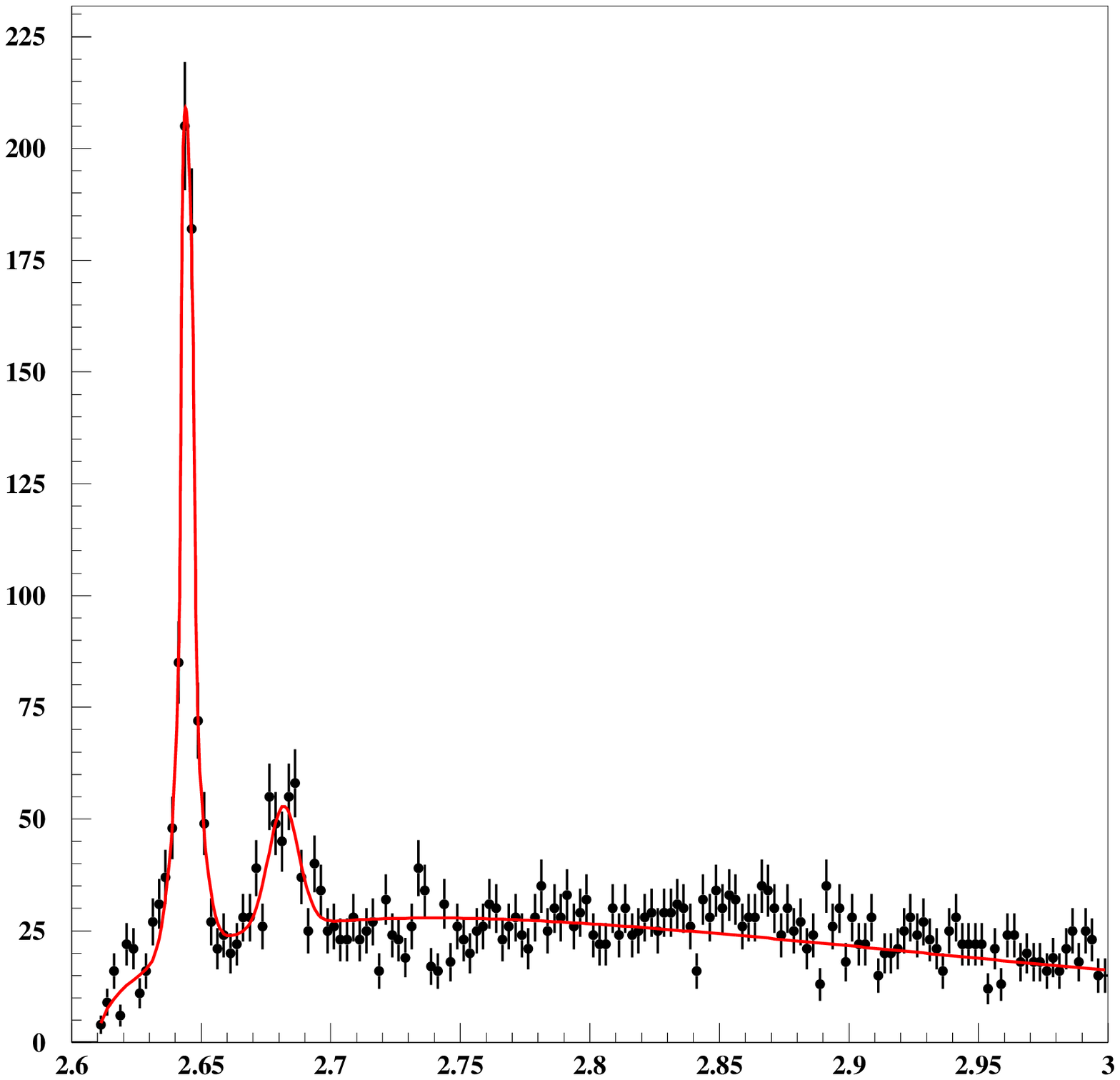,width=\linewidth}
\end{minipage}
\begin{minipage}[b]{.33\linewidth}
\centering\epsfig{figure=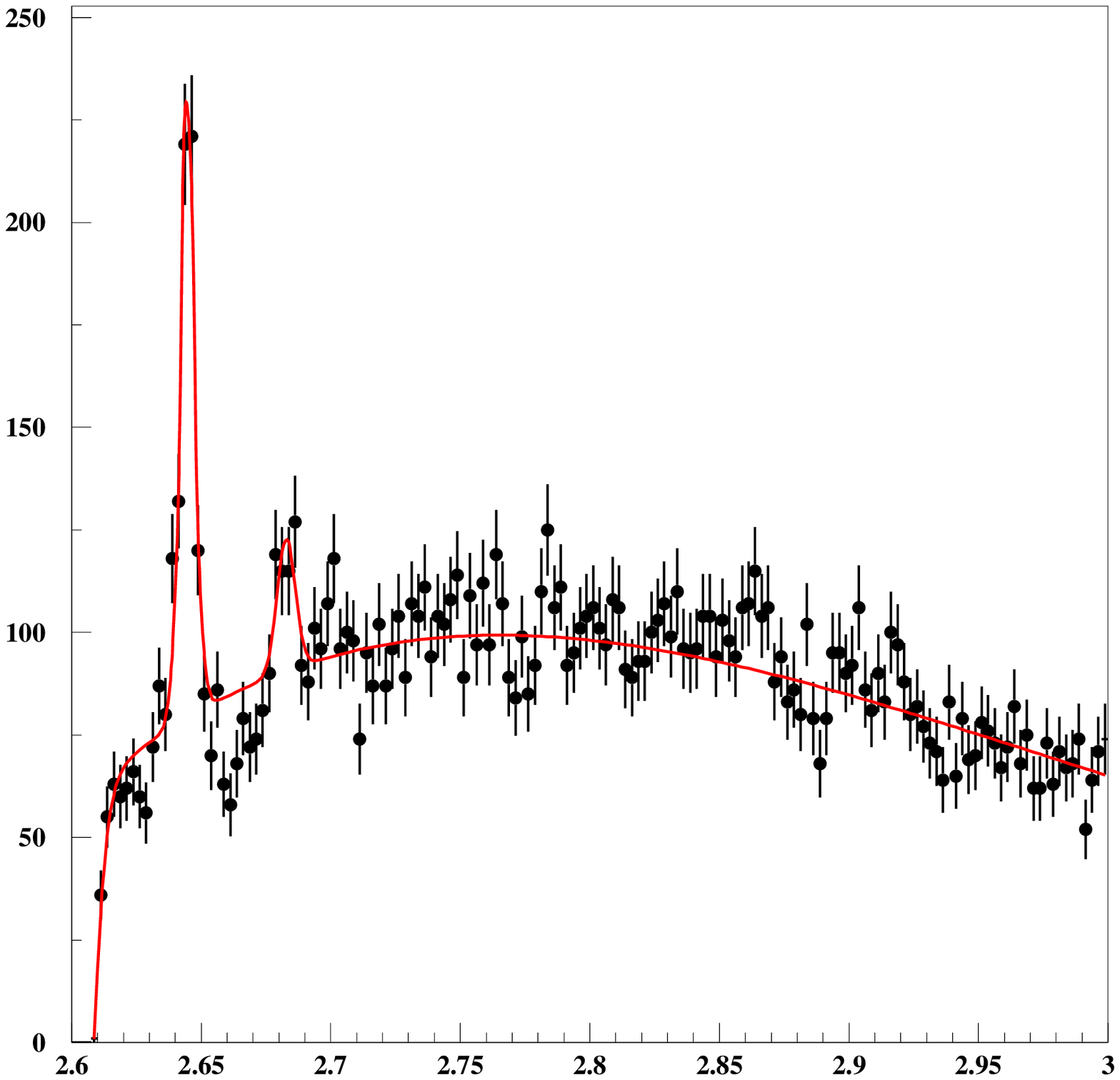,width=\linewidth}
\end{minipage}
\end{picture}
\caption
{Invariant mass distribution of the selected $\Xi_c\pi$ combinations (points with error bars)
together with the fit (solid curve) for  $\Xi_c^+\to \Xi^-\pi^+\pi^+$ (left),
 $\Xi_c^0\to \Xi^-\pi^+$ (middle) and $\Xi_c^0\to \Lambda K^- \pi^+$ (right plot).
The signal of the $\Xi_c(2645)$ (the second maximum, explained in the text) 
is described by a double (single)
 Gaussian, respectively. The background's shape is fixed by the $\Xi_c\pi$ spectra corresponding
to the $\Xi_c$ sidebands.
\label{XC2645}}
\end{figure}

Using the data sample of  357 fb$^{-1}$, the Belle collaboration reported at this 
Conference a precise measurement of the 
$\Xi_c(2645)^+$ and $\Xi_c(2645)^0$ baryon masses. The state $\Xi_c(2645)^0$ is observed
in the decay to $\Xi_c^+\pi^-$ ($\Xi_c^+\to\Xi^-\pi^+\pi^+$). The charged baryon
$\Xi_c(2645)^+$ is reconstructed in the decay to $\Xi_c^0\pi^+$ with the subsequent
decays of the $\Xi_c^0$ to $\Xi^-\pi^+$ or $\Lambda K^-\pi^+$ (Fig.~\ref{XC2645}).
In total, 1500 events of the above three decay modes are reconstructed. In addition, a second less
pronounced and broader maximum is observed in all plots of Fig.~\ref{XC2645} in the mass region
between 2.66 and 2.7 GeV/c$^2$. According to Monte Carlo studies, this peak is due to the
reflection from the decay chain:
$\Xi_c(2790)\to \Xi_c^{\prime}(2579)\pi, \Xi_c^{\prime}\to\Xi_c\gamma$, where the photon
is not observed.

The preliminary results of mass determination are
\begin{equation} 
m_{\Xi_c(2645)^+}  = (2644.7\pm 0.4\,{\rm (stat. \oplus syst.)}\pm 0.4)\,{\rm MeV}/c^2 
\label{XC2645P}
\end{equation} 
\begin{equation} 
m_{\Xi_c(2645)^0}  = (2643.1\pm 0.6\,{\rm (stat. \oplus syst.)}\pm 0.4)\,{\rm MeV}/c^2. 
\label{XC2645Z}
\end{equation} 
Here again the first error represents the combined statistical and systematic uncertainty,
and the second is the uncertainty due to possible biasses in the mass scale.
The $\Xi_c(2645)^+ - \Xi_c(2645)^0$ mass splitting is
\begin{equation} 
m_{\Xi_c(2645)^+} - m_{\Xi_c(2645)^0} = (1.6\pm 0.7)\,{\rm MeV}/c^2. 
\end{equation} 

\end{document}